\documentstyle[preprint,prl,aps,epsfig,rotating]{revtex}

\begin{document}
\draft

\title{Resonant formation of $d\mu t$ molecules in
deuterium:\\ an atomic beam measurement of muon catalyzed $dt$
fusion}

\author{M.~C.~Fujiwara,$^{1,2}$\cite{pread} A.~Adamczak,$^3$
J.~M.~Bailey,$^4$ G.~A.~Beer,$^5$ J.~L.~Beveridge,$^2$
M.~P.~Faifman,$^6$ T.~M.~Huber,$^7$ P.~Kammel,$^8$
S.~K.~Kim,$^{9}$ P.~E.~Knowles,$^5$\cite{preadPEK}
A.~R.~Kunselman,$^{10}$ M. Maier,$^{5}$ V.~E.~Markushin,$^{11}$
G.~M.~Marshall,$^2$ C.~J.~Martoff,$^{12}$ G.~R.~Mason,$^5$
F.~Mulhauser,$^2$\cite{preadPEK} A.~Olin,$^{2,5}$
C.~Petitjean,$^{11}$ T.~A.~Porcelli,$^5$\cite{preadTAP}
J.~Wozniak,$^{13}$ and J.~Zmeskal$^{14}$ \\ (TRIUMF Muonic
Hydrogen Collaboration)\\}

\address{$^1$Department of Physics and Astronomy,
 University of British Columbia, Vancouver, BC, Canada V6T 2A6\\
 $^2$TRIUMF, Vancouver, Canada, V6T 2A3\\
 $^3$Institute of Nuclear Physics, 31-342 Krakow, Poland\\
 $^4$Chester Technology, Chester CH4 7QH, England, UK\\
 $^5$Department of Physics and Astronomy, University of Victoria,
 Victoria, BC, Canada V8W 2Y2\\
 $^6$Russian Research Center, Kurchatov Institute, Moscow 123182,
 Russia\\
 $^7$Department of Physics, Gustavus Adolphus College, St. Peter, MN
 56082\\
 $^8$Department of Physics and Lawrence Berkeley National Laboratory,
 University of California, Berkeley, CA 94720\\
 $^{9}$Department of Physics, Jeonbuk National University, Jeonju
 City 560-756, S. Korea\\
 $^{10}$Department of Physics and Astronomy,
 University of Wyoming, Laramie, WY 82071-3905\\
 $^{11}$Paul Scherrer Institute, CH-5232 Villigen, Switzerland\\
 $^{12}$Department of Physics, Temple University, Philadelphia, PA
 19122\\
 $^{13}$Faculty of Physics and Nuclear Techniques,
 University of Mining and Metallurgy, 30-059 Krakow, Poland\\
 $^{14}$Institute for Medium Energy Physics, Austrian Academy of
 Sciences, A-1090 Vienna, Austria\\}

\date{\today}
\maketitle
\begin{abstract}

Resonant formation of $d\mu t$ molecules in collisions of muonic
tritium ($\mu t$) on D$_2$ was investigated using a beam of $\mu
t$ atoms, demonstrating a new direct approach in muon catalyzed
fusion studies. Strong epithermal resonances in $d\mu t$ formation
were directly revealed for the first time. From the time-of-flight
analysis of $2036\pm 116$ $dt$ fusion events, a formation rate
consistent with $0.73\pm (0.16)_{meas} \pm (0.09)_{model}$ times
the theoretical prediction was obtained. For the largest peak at a
resonance energy of $0.423 \pm 0.037$ eV, this corresponds to a
rate of $(7.1 \pm 1.8) \times 10^9$ s$^{-1}$, more than an order
of magnitude larger than those at low energies.

\end{abstract}
% Positronium, muonium, muonic molecules and atoms
% few body systems
% fusion reactions

Reactions of muonic hydrogen atoms and molecules present a
sensitive testing ground for few-body theories involving strong,
weak, and electromagnetic interactions. Among such reactions, muon
catalyzed fusion ($\mu$CF) in a deuterium--tritium mixture has
attracted particular interest, where one muon can catalyze more
than 100 nuclear fusions between deuteron and triton ($d+t
\rightarrow \alpha +n $)~\cite{fusions,review}.
A key step in high-yield $\mu$CF is resonant
formation~\cite{vesman} of the $d\mu t$ molecule
 \begin{equation}
    \mu t + D_2 \rightarrow [(d\mu t)_{11}dee]_{K\nu},
    \label{eq:mol-form}
 \end{equation}
where the collision energy, and the energy released upon formation
of $d\mu t$ in the loosely bound state ($Jv=11$), are absorbed in
the ro-vibrational ($K\nu$) excitation of the molecular complex
$[(d\mu t)dee]$, a hydrogen-like molecule with $(d\mu t)^+$
playing the role of one of the nuclei. Because of its limiting
rate $\lambda _{d\mu t}$, the process has been considered one of
the major bottlenecks in achieving high efficiency in $\mu$CF.
However, theoretical calculations \cite{faifm91,petrov} predict
strongly enhanced resonances for reaction~(\ref{eq:mol-form}) at
$\mu t$ kinetic energies of order 1 eV (epithermal resonances or
ER), yet the experimental information is scarce thus far. It is
the purpose of this letter to report a direct confirmation of the
predicted ER and a determination of their energies, enabled for
the first time with the atomic beam approach~\cite{marsh92}.

In conventional $\mu$CF experiments with homogeneous mixtures of
hydrogen isotopes~\cite{fusions,review}, the extraction of
$\lambda _{d\mu t}$ and its energy dependence relies on a kinetic
model describing a complex chain of reactions, which includes
processes that are not well understood, and assumptions that have
been recently challenged (see Ref.~\cite{q1s}). The collision
energy for reaction~(1) in the equilibrium states is given by the
target thermal energy, but difficulties in tritium handling have
so far prevented realization of high temperature targets capable
of thermally accessing ER.
Although transient phenomenagave first evidence~\cite{cohen85} and
subsequent insight~\cite{jeitl95} into the epithermal effects in
conventional targets, where a small fraction of $d\mu t$ formation
may occur before the thermalization, no quantitative measurement
of $\lambda _{d\mu t}$ for ER or its energy $E_{res}$ has been
obtained due primarily to unknown $\mu t$ initial conditions.

The atomic beam method described here provides direct access to ER
because of the available $\mu t$ beam energy (0.1--10 eV). A
single $d\mu t$ formation can be studied, on an event-by-event
basis, isolated in time and space from other processes. With the
$\mu t$ time of flight (TOF) between the separated layers
providing a measure of the collision energy, $E_{res}$ can be
determined in a direct manner. Furthermore, manipulation of
heterogeneous multi-layers allows us to study and control,
individually, important processes such as $\mu$ transfer, $\mu t$
emission, and moderation.

The experiment was performed at the M20B channel at TRIUMF.
Details of the measurement and analysis may be found in
Ref.~\cite{fujiw99b}. The apparatus~\cite{knowl96,nim_thick} is
illustrated in Fig.~1. Target layers (Fig.~2(a)) were prepared by
rapidly freezing a hydrogen isotope or mixture onto the gold
foils, held at 3.5 K in an ultra-high vacuum of order 10$^{-9}$
torr or better.  The layer thickness and uniformity were
characterized off-line via energy loss of $\alpha$
particles~\cite{nim_thick}, and the effective thickness was then
derived using a separate measurement~\cite{fujiw99} of the muon
beam profile (about 30~mm diameter).

A beam of $5 \times 10^{3} \mu^-\ \rm{s}^{-1}$ of momentum $p$=27
MeV/$c$ and momentum spread $\Delta p/p$=5.5\% (FWHM), defined by
a 250 $\mu$m scintillator of diameter 48~mm (T1), entered the
target vacuum. It was degraded mainly by a 51 $\mu$m gold foil
upstream (US) from a TOF region. A fraction $S_F$ of the $\mu$
stopped in a tritium-doped hydrogen layer (an emission layer) of
3.43$\pm$0.18 mg$\cdot$cm$^{-2}$ frozen to the US foil; most of
these initially formed $\mu p$. The transfer $\mu p \rightarrow
\mu t$ took place in a time of typically 100 ns~\cite{mulha96},
creating $\mu t$ with recoil energy of 45 eV due to the reduced
mass difference. Because of the Ramsauer-Townsend effect in $\mu t
+ p$ scattering,
$\mu t$ atoms are emitted with energies near 10 eV into an
adjacent layer~\cite{fujiw99}. A D$_2$ moderation layer
(96.0$\pm$5.0 $\mu$g$\cdot$cm$^{-2}$) efficiently reduced the $\mu
t$ beam energy to $\sim$1 eV  via elastic scattering in order to
better match ER energies.
The $\mu t$, after a flight time of a few $\mu$s, reached the
D$_2$ reaction layer downstream (DS) from the TOF region
(17.9$\pm$0.5~mm from the US layers) and formed $[(d\mu t)dee]$
from which fusion can occur to produce $\alpha$ and $n$. The time
between the $\mu$ entrance signal and the detection of a fusion
$\alpha$ (which we call {\it fusion time}) is dominated by $\mu t$
TOF and provides information on molecular formation energy, as
long as the energy loss ($\Delta E_{\mu t}$) of $\mu t$, due to
elastic scattering before the formation in the DS D$_2$, is small.
A thin DS layer of 21.2$\pm$1.4 $\mu$g$\cdot$cm$^{-2}$ was chosen
to minimize $\Delta E_{\mu t}$ so as not to obscure the
time-energy correlation. The detailed theoretical description of
our method is given in Ref.~\cite{marku96}.

Two series of data, with emission layer tritium concentrations
$c_t =0.1\%$ (Run 1) and $c_t=0.2\%$ (Run 2), were analyzed
separately. Deuterium concentrations in H$_2$ were typically less
than 2~ppm. Run 1 had a 1.77$\pm$0.12 mg$\cdot$cm$^{-2}$ H$_2$
substrate beneath the D$_2$ DS reaction layer, while for Run 2 the
D$_2$ reaction layer was deposited directly on the DS gold foil.
The data were normalized to the corrected number of incident
muons, $N_\mu$, which takes into account the data acquisition dead
time ($\sim$20\%) and pile-up of incident muons in the 10 $\mu s$
gate ($\sim$5\%).
The signal and background in the Si detectors for Run 1 are shown
in Fig.~2(b) with a delayed time cut selecting DS events, giving a
S/B ratio of about 2:1. The background, mainly due to $\mu$ decay
and capture related processes (few fusion events from the US D$_2$
moderator pass the delayed time cut because there is no TOF
delay), can be accurately determined with the beam method from
runs without the DS reaction layer, in which only DS fusion is
turned off while other processes are not affected. Potential
background from protons from $\mu$-induced $dd$ fusion, arising
for example from a recycled $\mu$ following $dt$ fusion, were
estimated to give a $2.4 \pm 0.9$\% correction in DS fusion by
looking at $\alpha$--$p$ correlations in the Si detectors, using
the much larger data sample from fusion in the US moderation
layer. A small contamination ($\lesssim 5$ ppm) of nitrogen in the
target layer was estimated to reduce the fusion yield by about
2\%.
Any further residual effects ({\it e.g.,} due to inaccuracy in the
$N_\mu$ scaler, or time zero shifts) were carefully investigated
and were conservatively reflected in the final errors.
With totals for  $N_\mu$ of $6.02\times 10^{8}$ and $2.82 \times
10^8$ for the data and background runs, respectively, we have
observed $2036\pm116$ DS fusion events, for Runs 1 and 2 combined.

An absolute measurement of the fusion yield required determination
of several factors including stopping fraction $S_F$, Si detector
solid angle $\Omega _{Si}$, and energy and time cut acceptances,
$\epsilon _E$ and $\epsilon _T$. Our value of $S_F = 0.299\pm
0.015$ was based on fits of the decay electron time spectra
recorded by electron and neutron counters (in the charged mode),
where $\mu$ stopped in H$_2$ exhibited a characteristic $\sim$2
$\mu$s lifetime.
The absolute efficiency of the counters was determined from the
detection of delayed electrons following the observation of
fusion~\cite{fujiw99b}. $S_F$ then could be derived
model-independently from the absolute amplitude at time zero of
the 2 $\mu$s decay electron component, normalized to $N_\mu$. An
independent and consistent estimate of $S_F$ was obtained from a
GEANT~\cite{geant} beam and decay simulation, which reproduced our
measured range curve~\cite{foot}.

The energy cut efficiency and possible existence of tails in the
$\alpha$ energy distribution in the Si spectrum (Fig. 2(b)) were
investigated by changing the cut width, as well as by simulating
the energy loss of $\alpha$ particles in the DS D$_2$
layer~\cite{pepper}, giving $\epsilon _E = 0.978 \pm 0.064$ for
the applied cut of $3.1<E_{\alpha}<3.7$ MeV. $\Omega _{Si}$ was
determined to be $2\times (2.46\pm 0.10)$\% from similar
calculations, in which the spatial distribution of fusion events
was constrained by electron imaging of muon decay. Effects of
possible beam shifts~\cite{foot2} as well as small geometrical
inaccuracy are reflected in the errors. After combining all the
factors, including a small time cut correction, the absolute
normalization was determined to the relative precisions of 9.5\%
and 7.7\% for Runs 1 and 2, which together with statistical
uncertainties gave the DS fusion yield per stopped $\mu$ of
$(5.04\pm 0.50)\times 10^{-4}/\mu$ and $(4.27\pm 0.54)\times
10^{-4}/\mu $ for Runs 1 and 2, respectively.

The results were used to test theoretical predictions as follows.
The formation rate $\lambda _{d\mu t}^{th}$ from
Ref.~\cite{faifm91}, corrected for the Doppler broadening due to
the target molecule motion at 3 K (without many-body effects), was
used as a standard input, with low energy rates further modified
to reflect possible sub-threshold effects~\cite{petrov}. The
experimental TOF fusion distributions were fitted with a series of
Monte Carlo (MC) simulated spectra for which the input $\lambda
_{d\mu t}$ and $E_{res}$ were varied as $\lambda _{d\mu t} = {\cal
S}_{\lambda} \lambda _{d\mu t}^{th}$ and $E_{res} = {\cal S}_E
E_{res}^{th}$. The measured absolute normalization and its
estimated uncertainty were used to constrain the fit. The best
values for ${\cal S}_{\lambda}$ and ${\cal S}_E$, which were taken
to be energy- and rate-independent, respectively, were extracted
from the $\chi^2$ distribution. The fusion probability $W_f$ from
Ref.~\cite{faifm91} was kept fixed during the fit.
A detailed simulation code was developed for the
analysis~\cite{huber}. Input cross sections for muonic processes
were based on theoretical values in Refs.~\cite{atlas,faifm89},
but our multi-layered target allowed independent tests of
important reactions including $\mu p \rightarrow \mu t$ transfer
and $p\mu p$ formation~\cite{mulha96}, and $\mu t + p$ and $\mu t
+ d$ collisions~\cite{fujiw99b,fujiw99}. In addition, independent
simulation codes~\cite{marku96,wozni96} were used to check the
consistency of some of the key processes.
We stress the importance of explicitly treating the resonant
scattering in the calculations; it can be shown~\cite{fujiw99b}
that the use of the renormalized effective rate $\tilde{\lambda}
_{d\mu t}\sim W_f \lambda _{d\mu t}$ as in Ref.~\cite{marku96}
would result in the overestimate of the fusion yield by as much as
a factor of two, a fact which resolves an earlier discrepancy
reported in Ref.~\cite{marsh92}. The effect of the energy
distribution of resonantly scattered $\mu t$ was investigated
within the kinematically possible range.
Solid state effects in thermalization and formation~\cite{solid},
%*** which are important mostly lowest at energies, were not explicitly
which are important mostly at lowest energies, were not explicitly
included in the simulations, but their influence at ER energies is
expected to be small. The effects were estimated by appropriately
modifying the cross sections. Table~\ref{tab1} summarizes our
evaluation of systematic uncertainties in the MC modelling.

Figure~\ref{fig:mc} shows a comparison of the calculated MC
spectrum with the data for Run 1. Also plotted are the simulated
contributions from the time--energy correlated events ({\it i.e}.\
with small $\Delta E_{\mu t}$), exhibiting more clearly the
resonance structures due to different $\nu$ states. From the fits,
we obtained ${\cal S}_\lambda= 0.88 \pm 0.11$ (Run 1) and $0.55\pm
0.12$ (Run 2) with $\chi ^2/\rm{dof}$ ($\rm{dof}= 50$) of 0.96 and
1.27 respectively, and ${\cal S}_E=0.928 \pm 0.040$ (Run 1) and
$0.994\pm 0.087$ (Run 2) with similar fit quality. The fit
uncertainties reflect both statistical and normalization errors.
An apparent deviation at the 2$\sigma$ level in $S_\lambda$
between Runs 1 and 2 may indicate some unaccounted-for systematic
uncertainties (due possibly to slightly different experimental
conditions as described above), hence the measurement error was
increased accordingly so that two data set give $\chi ^2$ of 1
~\cite{PDG}. When the known MC modelling error (Table~\ref{tab1})
is included, the combined final result is ${\cal S}_\lambda = 0.73
\pm (0.16)_{meas} \pm (0.09)_{model}$. As for the resonance energy
measurements, the weighted average of two runs gave ${\cal
S}_E=0.940 \pm (0.036)_{meas} \pm (0.080)_{model}$, where the
modelling error is mainly due to uncertainties in the $\mu t + d$
elastic scattering process and the $\mu t$ TOF drift distance.

These scaling results correspond, in the theory~\cite{faifm91}, to
a peak formation rate for the main resonance ($\nu$=3) of $(7.1\pm
1.8)\times 10^9$ s$^{-1}$ at $E_{res}=0.423\pm 0.037$ eV for $\mu
t^{F=1}$, and $(6.5\pm 1.6)\times 10^9$ s$^{-1}$ at $0.508\pm
0.047$ eV for $\mu t^{F=0}$, where $F$ refers to the $\mu t$
hyperfine state (the rates are normalized to liquid hydrogen
density, $4.25\times 10^{22}$ cm$^{-3}$). The obtained rates are
more than an order of magnitude larger than those at low
energies~\cite{fusions}, experimentally demonstrating the prospect
for high cycling $\mu$CF in a high temperature target of several
thousand degrees.

Our TOF sensitivity for the energy dependence allows us to clearly
reject, for example, a constant $\lambda _{d\mu t}$, establishing
the existence of resonant structure in epithermal $d\mu t$
formation. If one assumes the energy level spectrum of the $[(d\mu
t)dee]$ molecule, our results for $E_{res}$ imply sensitivity to
the binding energy of the $(d\mu t)_{11}$ state with an accuracy
of the order of the vacuum polarization corrections.

We have recently collected data for resonant $d\mu t$ formation in
$\mu t + \rm{HD}$ collisions, for which the predicted resonances
are even stronger~\cite{faifm91,petrov}. The results of that
measurement will be reported in a future
publication~\cite{porce00}.

We wish to thank C. Ballard, J. L. Douglas, K. W. Hoyle, R.
Jacot-Guillarmod, and N. P. Kherani for their valuable help. The
support of TRIUMF and its staff is gratefully acknowledged. M.C.F.
would like to thank D. F. Measday for many helpful comments. This
work was supported in part by NSERC (Canada), DOE and NSF (USA),
and NATO Linkage Grant LG 930162.

\begin{table}
  \begin{center}
    \leavevmode
    \caption{Estimated effects on rate scaling parameter
     ${\cal S}_\lambda$ by the systematic uncertainties
     in the MC modelling.}
    \begin{tabular}{lr}
MC error source & $\Delta {\cal S}_\lambda/{\cal S}_\lambda$ \%
 \\ \hline
 $\mu$ beam size & 1.2 \\
 Nonuniform $\mu$ stopping (GEANT) & 1.8 \\
 $\mu t$ TOF drift distance & 2.6 \\
 $\mu p \rightarrow \mu t$ transfer, $p\mu p$ formation & 5.6 \\
 $\mu t + d$, $\mu t + t$ scattering, layer thickness & 5.7 \\
 $\mu t + p$ RT minimum energy  & 1.1 \\
 Resonance Doppler widths in solid & 8.0 \\
 Solid state and low energy processes & 5.0 \\
 (subthreshold resonances, slow thermalization, & \\
 $\mu t$ energy after resonant scattering) & \\ \hline
 Total (in quadrature)  &  12.9 \\
    \end{tabular}
    \label{tab1}
  \end{center}
\end{table}

\begin{figure}
 \begin{center}
 \epsfig{file=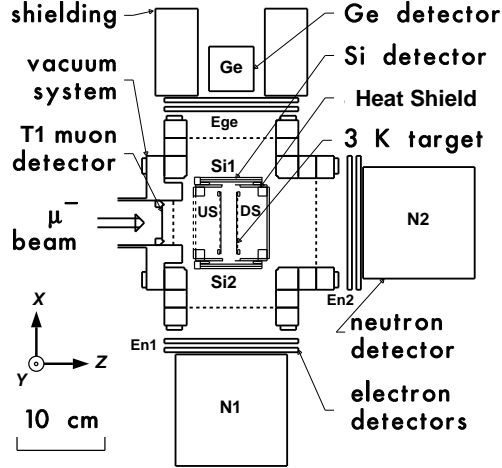,width=0.40\textwidth}
  \end{center}
 \caption{Top view of the apparatus. Si detectors were placed
 in vacuum viewing target layers without a window, enabling the
 high resolution detection of
 fusion $\alpha$ particles (Fig.~2(b)).
 The Ge detector monitored target impurities via muonic
 X-rays, while plastic scintillators (En1, En2, Ege) were used
 both to detect muon decay electrons and to veto charged particles
 for Ge and neutron (N1, N2) detectors.}
 \label{fig:app}
 \end{figure}

\begin{figure}
 \begin{center}
 \epsfig{file=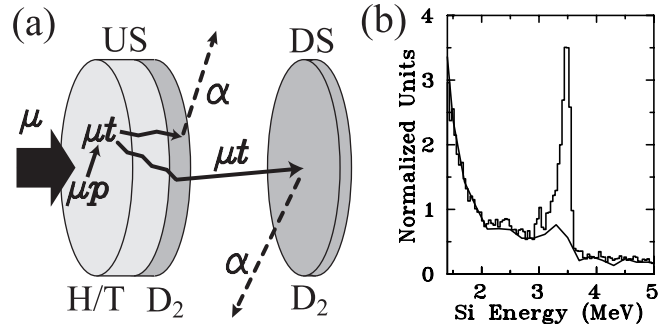,width=8.6cm}
 \end{center}
\caption{(a) Schematic of the target consisting of emission,
moderation, and reaction layers (not to scale). (b) Si energy
spectra showing signal (histogram) and background (line) with a
time cut of $t>1.5$ $\mu$s selecting DS events delayed by $\mu t$
TOF.}
 \label{fig:siE}
 \end{figure}

 \begin{figure}
 \begin{center}
 \epsfig{file=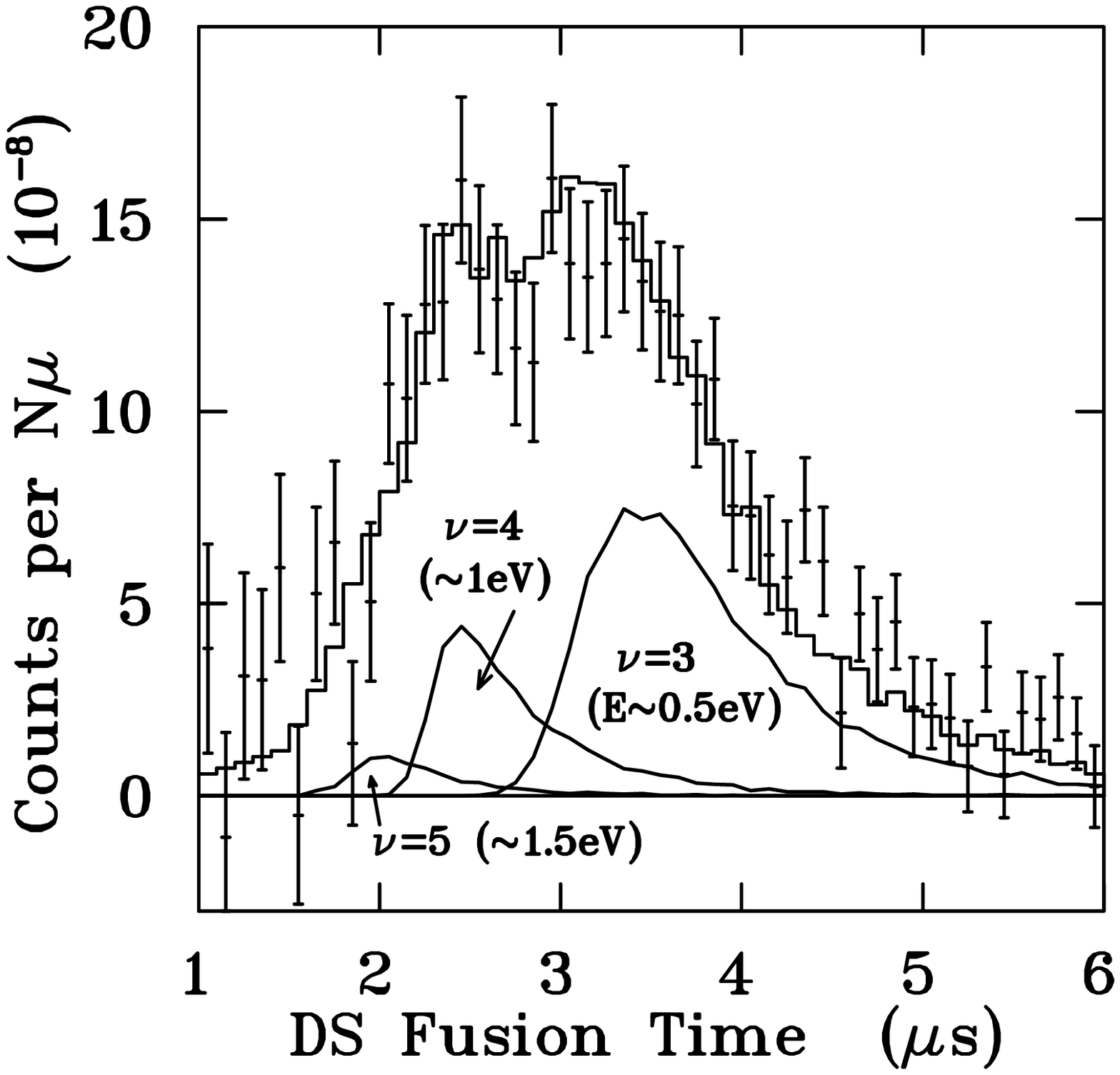,width=0.35\textwidth}
 \end{center}
 \caption{Time-of-flight fusion spectrum (error bars) and
 simulation spectrum (histogram). Also plotted are simulated
 contributions from different resonance peaks given by
 time--energy correlated events with $\Delta E_{\mu t} < 0.15$
 eV. Note that angular dispersion of the $\mu t$ beam also
 contributes to the peak widths.}
 \label{fig:mc}
 \end{figure}


\begin{references}
\bibitem{foot} We note that the commonly used method of estimating
$S_F$ from relative amplitudes of a multi-lifetime fit (H$_2$
versus Au) in the decay electron time spectrum gave an
inconsistent result, which we attribute to a low energy background
related to $\mu$ capture on Au.
\end{references}
\end{document}